\documentclass[conference, twocolumn, final]{IEEEtran}
\IEEEoverridecommandlockouts
\usepackage{algorithmic}
\usepackage{textcomp}
\usepackage{cite}

\usepackage{amsmath,graphicx, amsthm}
\usepackage[dvipsnames]{xcolor}
\usepackage{amssymb}
\usepackage{amsfonts, dsfont}
\usepackage{booktabs}
\usepackage{comment}
\usepackage{enumerate}
\usepackage{hyperref}
\usepackage{pgfplots}
\usepackage{multirow}
\usepackage{mathrsfs}
\usepackage{optidef}
\usepackage{float}
\usepackage{pgfplotstable} 

\DeclareMathOperator*{\argmax}{arg\,max}
\newtheorem{lemma}{Lemma}

\pgfplotsset{compat=newest,
    width=3.5cm,
    height=3cm,
    scale only axis=true,
    grid style={dashed},
    max space between ticks=25pt,
    try min ticks=5,
    scaled x ticks=false,
    every axis plot/.append style={thick},
    tick style={black, thick},
    legend style={font=\footnotesize},
    legend image code/.code={
    \draw[mark repeat=2,mark phase=2]
        plot coordinates {
        (0cm,0cm)
        (0.1cm,0cm)        
        (0.2cm,0cm)         
        };%
    }
}
\tikzset{
    semithick/.style={line width=1pt},
}
\usepgfplotslibrary{groupplots}
\usepgfplotslibrary{dateplot}
\usepgfplotslibrary{fillbetween}

\makeatletter
\newcommand*\titleheader[1]{\gdef\@titleheader{#1}}
\AtBeginDocument{%
  \let\st@red@title\@title
  \def\@title{%
    \bgroup\footnotesize\centering\@titleheader\par\egroup
    \vskip1.5em\st@red@title}
}
\makeatother

\title{Sustainable Edge Intelligence Through Energy-Aware Early Exiting\thanks{This work received funding from the EU Horizon 2020 Marie Sk\l odowska Curie ITN Greenedge (GA. No. 953775), and UKRI through project SONATA (EPSRC-EP/W035960/1). For the purpose of open access, the authors have applied a Creative Commons Attribution (CC BY) license to any Author Accepted Manuscript version arising from this submission.}
}
\titleheader{This paper has been accepted for presentation at IEEE International Workshop on Machine Learning for Signal Processing (MLSP) 2023. ©2023 IEEE. 
}
\author{\IEEEauthorblockN{ Marcello Bullo\IEEEauthorrefmark{1}\IEEEauthorrefmark{2}, Seifallah Jardak\IEEEauthorrefmark{1}, Pietro Carnelli\IEEEauthorrefmark{1} and Deniz G\"und\"uz\IEEEauthorrefmark{2}}\\
\IEEEauthorblockA{\IEEEauthorrefmark{1}Bristol Research and Innovation Laboratory (BRIL), Toshiba Europe Ltd., Bristol, UK}
\IEEEauthorblockA{\IEEEauthorrefmark{2}Department of Electrical and Electronic Engineering, Imperial College London, UK}
\small{Emails:\{marcello.bullo, seifallah.jardak, pietro.carnelli\}@toshiba-bril.com, \{m.bullo21, d.gunduz\}@imperial.ac.uk }
}

\begin{document}

\maketitle

\begin{abstract}
Deep learning (DL) models have emerged as a promising solution for the Internet of Things (IoT). However, due to their computational complexity, DL models consume significant amounts of energy, which can rapidly drain the battery and compromise the performance of IoT devices. For sustainable operation, we consider an edge device with a rechargeable battery and energy harvesting (EH) capabilities. In addition to the stochastic nature of the ambient energy source, the harvesting rate is often insufficient to meet inference energy requirements, causing drastic performance degradation in energy-agnostic devices.
To mitigate this problem, we propose energy-adaptive dynamic early exiting (EE) to enable efficient and accurate inference in an EH edge intelligence system. Our approach derives an energy-aware EE policy that determines the optimal amount of computational processing on a per-sample basis. The proposed policy balances energy consumption to match the limited incoming energy and achieves continuous availability. Numerical results show that accuracy and service rate are improved up to 25\% and 35\%, respectively, compared to an energy-agnostic policy.
\end{abstract}

\section{Introduction}

The emergence of deep learning (DL) algorithms has enabled data-driven services for Internet of Things (IoT) applications at the network edge. However, limitations in energy resources and computing power constrain the complexity of DL architectures and restrict the data processing algorithms that can be deployed on edge devices. In this regard, energy efficiency is becoming a critical design factor for DL applications \cite{AIenergy}. Although their energy footprint has considerably reduced thanks to architectural and computational improvements \cite{edgeDev}, the energy consumption of inference tasks represents a dominant concern for the future of edge intelligence, particularly when the same model is used repeatedly on many samples.

A promising solution for the sustainability of intelligent IoT applications is the energy harvesting (EH) technology~\cite{green_EH}. While EH promises perpetual operation, the uncertainty of the available ambient energy and the limited capacity of the rechargeable battery represent major challenges in the design and operation of EH devices \cite{sustIoT}. While minimizing the energy consumption is the current goal in approaches to sustainable edge intelligence, the primary objective with EH technology is the intelligent management of available energy for prolonged operation.

The paper studies sustainable edge intelligence with an EH device and a rechargeable battery. We consider a deep neural network (DNN) model that performs resource- and input-adaptive inference. We adopt early exiting (EE) \cite{SCEE} as an adaptive energy-aware inference strategy. A time-correlated process models the stochastic availability of the harvested energy resources, and a finite discrete buffer simulates the limited-capacity battery.
The problem is presented as a discrete Markov decision process (MDP) where, at each decision slot, the controller chooses between EE and full computation based on the input sample, the EH process, and the battery level. The overall inference accuracy depends not only on the DNN architecture but also on the availability of resources. Thus, the controller's decision-making process becomes critical in balancing accuracy and energy sustainability. 
Our contributions are summarized as follows:
\begin{itemize}
    \item We provide a theoretical analysis to optimize the controller's behaviour in the steady-state regime, yielding an optimal energy-aware policy that adapts to the limited and stochastic nature of the energy source.
    \item We design a feasible causal controller that imitates the behaviour of the derived optimal non-causal controller using a proposed Bayesian predictor.
    \item We validate our theoretical findings with numerical results for a sustainable image classification application.
\end{itemize}
To the best of our knowledge, this is the first paper that studies the impact of stochastic energy availability on the edge inference accuracy, and adapts the inference strategy using EE, to judiciously utilize the limited energy resources.

\section{System Model}\label{sec:system_model}
\begin{figure}[t]
    \centerline{\includegraphics[scale=.58]{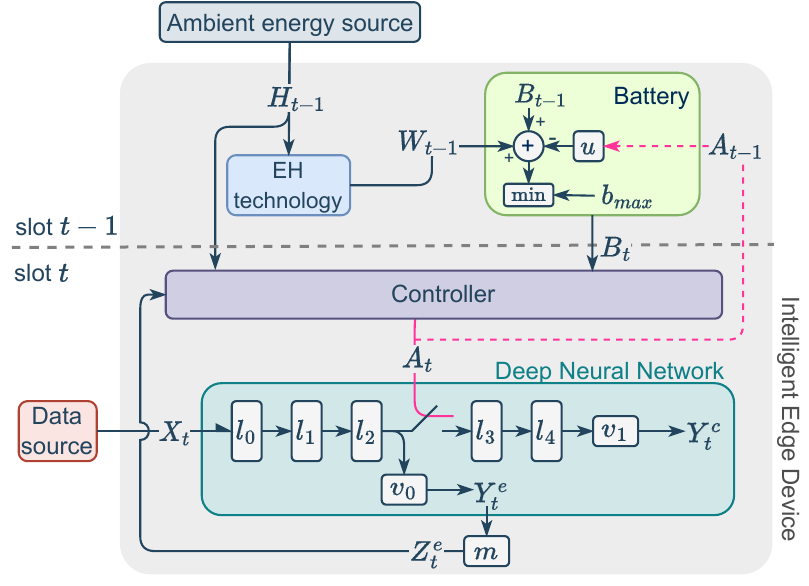}}
    \caption{Block diagram of the intelligent edge device with EH capabilities and rechargeable battery.}
    \label{fig:EHdevice_schema}
\end{figure}

In this section, we provide a mathematical model of the edge device considered for this work, and we characterize its interactions with the ambient energy and data processes. The block diagram of the overall system is depicted in Fig.~\ref{fig:EHdevice_schema}.

\noindent The EH module gathers energy from the environment and converts it to a ready-to-store form. The accumulated energy is stored in the battery to power a DNN model implementing EE. The controller is responsible for managing the computational processing of data in the DNN to balance between energy consumption and performance.

The \emph{ambient energy source process} $\{H_t\}$ is modeled as a two-state discrete-time Markov process with state space $\{G,\beta\}$. The state $G$ represents a \emph{good} harvesting condition, while $\beta$ represents a \emph{bad} one. The probability of remaining in state $G$ and in state $\beta$ are denoted by $p_G$ and $p_\beta$, respectively. 

The \emph{harvested energy arrival process} $\{W_t\}$ captures the amount of harvested energy at time $t$. In \emph{good} conditions, the system can harvest 0, 1, or 2 energy quanta with probability $\lambda_i = P[W_t=i|H_t=G], \text{for }i=0,1,2$. In \emph{bad} conditions, no energy is harvested, $P[W_t=0|H_t=\beta]=1$. The long-term average rate of the harvested energy can be expressed as $\varepsilon=(\lambda_1+2\lambda_2)P_G^{\infty}$, where $P_G^{\infty}$ is the steady-state probability of observing the ambient energy source process in state $G$.  In the special case where the energy source is limited but always available, i.e., no intermittency, the problem can be formulated by setting $p_G=1$, $p_\beta=0$, and the incoming energy quanta to the target energy rate. We denote the amount of energy quanta accumulated in the battery at time $t$ as $B_t$, where $B_t \in \mathcal{B} \triangleq \{0, 1, \dots , b_{max}\}$.

At each slot $t$, a data sample $X_t$ arrives at the device. Let $Y_t$ denote the correct label of $X_t$. We assume that the samples are independent and identically distributed (i.i.d.), and their labels are uniformly distributed over a finite set $\mathcal{Y}$. We define a DNN as $f_\theta(X_t)$, where $\theta\in\mathbb{R}^D$ represents the parameter vector. Without loss of generality, $f$ is modeled as the composition of $L$ differentiable operators $l_i,\ i=0,\dots,L-1$. Early exit classifiers $v_k,\,k=0,\dots,K-1$ are attached to the DNN structure in order to trade-off accuracy with reduced computations, which translates into lower energy consumption. We assume that the duration of a time slot is sufficient to run the whole DNN $f_\theta$ on the input.
In this paper, we will consider only one EE point, i.e., $K=1$. Hence, the controller chooses one of the three possible actions: (i) \emph{(d)iscard} the data sample without any inference, (ii) \emph{(e)xit}  the computation after the EE point, (iii) \emph{(c)ontinue} the computation until the end. The action space is defined as $\mathcal{A}=\{d,e,c\}$, and the action taken at time $t$ is $A_t$. Each action is associated with an energy cost which depends on the computation complexity of the DNN model. We denote the energy cost of $a\in\mathcal{A}$ as $u(a)$, where $0=u(d)<u(e)<u(c)$. The evolution in time  of the battery store process is governed by $A_t$ and $W_t$, and it is expressed by \mbox{$B_{t+1} = \min\{ B_t - u(A_t) + W_t, b_{max} \}$}.\\
\indent The goal of the controller is to choose actions to maximize the probability of correct inference over a long-term horizon. Initially, it can choose to discard or run the DNN up to the first exit point $v_0$. If the latter action is selected, the next decision can depend on the output of the DNN at this point, denoted by $Y^e_t$. Hence, the controller can select multiple actions sequentially at each time step.   To simplify the formulation, the action $d$ is taken only if the energy in the battery is not enough to run the DNN at all, i.e., $B_t<u(e)$. Otherwise, the controller decides between actions $e$ and $c$ based on $Y^e_t$.

\section{Problem Formulation}
The problem is modeled as an infinite-horizon sequential decision process.
At each decision time $t$, the system occupies state $S_t = (B_t, H_{t-1})\in \mathcal{S}$, and a side-information vector $Z_t\in\mathcal{Z}$ arrives according to a continuous distribution $\mu_Z$, $\forall t$. We consider $Z_t$ as a bounded-value vector computed as a function of the DNN outputs at time $t$. Note that at inference time, $Z_t$ is a deterministic function of the input $X_t$, thus i.i.d.. In each state, the controller chooses the action $a\in\mathcal{A}$, leveraging $Z_t$. To do so, a Borel-measurable function $g_s\in\mathcal{G}=\{g:\mathcal{Z}\to\mathcal{A}\}$, is defined $\forall s$, mapping the side information into actions. Each function $g_s$ induces a probability distribution over $\mathcal{A}$, that is $p_a = \mu_Z(g_s^{-1}(a))$, $a\in\mathcal{A}$, where $g_s^{-1}$ is the preimage of $g_s$. Thus, the controller chooses the appropriate $g$ in state $s\in\mathcal{S}$, a reward signal $r(s,g)$ is observed, and a state transition to $s'$ occurs according to $P(s'|s,g)$.

We formulate the problem as an MDP with side-information (MDP-SI) \cite{samuel}, defined as $\mathcal{M}=(\mathcal{S}, \mathcal{G}, P, r, \mathcal{Z}, \mu_Z)$. The controller maximizes the long-term inference accuracy of the DNN model, i.e.
\begin{equation}\label{eq:acc}
    \limsup_{T\to\infty} \mathbb{E}\Bigg[\frac{1}{T}\sum_{t=0}^{T-1}\mathds{1}_{\{\hat{Y}_t^{A_t}=Y_t\}}(S_t,A_t)\Bigg],
\end{equation}
where $\mathds{1}_{\{E\}}$ represents the indicator function for the event $E$, and $\hat{Y}^{A_t}_t$ denote the labels output when the action $A_t$ is decided. We assume that the reward is zero whenever $A_t=d$, i.e., $\mathds{1}_{\{\hat{Y}_t^{d}= Y_t\}}(S_t,d)=0, \forall t$. The above expectation is computed w.r.t. the MDP dynamics and $Z_t$, since $A_t = g_{S_t}(Z_t)$ is not deterministic. However, because $Z_t$ is assumed i.i.d., we can drop its statistical dependence on time and define the expected immediate reward as 
\begin{equation}\label{eq:exp_immediate_reward_ideal}
    r\big(S_t, g_{S_t}\big) = \mathbb{E}_{Z}\bigg[ \mathds{1}_{\{\hat{Y}_t^{g_{S_t}(Z)}=Y_t\}}\big(S_t,g_{S_t}(Z)\big)\bigg].
\end{equation}
Similarly, $ P(s'|s,g) = \mathbb{E}_{Z}[\text{Pr}(s'|s,g(Z))]$.

\section{System Optimization}
The objective of the controller is to find an action strategy that maximizes the long-term average reward. A policy is defined as a mapping from state to action, i.e., $\pi:\mathcal{S}\to\mathcal{G}_s$, with $\pi(s)=g_s$. Thus, the optimization goal is to identify the optimal policy $\pi^* = \argmax_{\pi\in\Pi} \rho_\pi$, where
\begin{equation}
    \rho_\pi = \limsup_{T\to\infty} \mathbb{E}_{\pi} \bigg[ \frac{1}{T}\sum_{t=0}^{T-1} r(S_t, \pi(S_t)) \bigg].
\end{equation}
For a unichain MDP with finite states and actions, the average reward optimality equation, $\forall s\in\mathcal{S}$, is
\begin{equation}\label{eq:optimality}
    \max_{g\in\mathcal{G}_s}\Big\{ r(s,g) - \rho^* + \sum_{s'\in\mathcal{S}}P(s'|s,g)V^*(s') - V^*(s)  \Big\}=0.
\end{equation}
Although $\mathcal{M}$ has a finite state space, the set of feasible choices in each state $s$, $\mathcal{G}_s$, is infinite. For \eqref{eq:optimality} to hold, $\mathcal{G}_s$ must be a compact space \cite{Puterman}. Therefore, we use an approximation of \eqref{eq:exp_immediate_reward_ideal} which allows us to work on a compact space. Let $Z^e_t = \lVert  Y^e_t\rVert_\infty$ denote the confidence of the model at \emph{early exit}, while $Z^c_t = \lVert  Y_t^c \rVert_\infty$ be the confidence computed at the final output. We employ $Z_t^e$ and $Z_t^c$ as an estimate of the true likelihood at early and final stages, respectively. Hence, we define the expected immediate reward as 
\begin{align}\label{eq:exp_immediate_reward}
    \begin{split}
    \hat{r}(S_t,g_{S_t})
    &=  \mathbb{E}_{\mathbf{Z}}\bigg[  Z^e\mathds{1}_{\{g_{S_t}^{-1}(e)\}}(\mathbf{Z}) + Z^c\mathds{1}_{\{g_{S_t}^{-1}(c)\}}(\mathbf{Z})\bigg],
    \end{split}
\end{align}
where $\mathbf{Z}=[Z^e, Z^c]$ is the side-information vector, $\mathbf{Z}\in\mathcal{Z}=\mathcal{Z}^e\times\mathcal{Z}^c$. In the following sections, we derive the optimal policy within this setting.

\subsection{Optimal Non-Causal Controller}
The expected immediate reward expressed in \eqref{eq:exp_immediate_reward} cannot be computed at the early exit stage, as it requires the knowledge of the future final prediction $\hat{Y}_t^{c}$. We consider an ideal setting where an optimal non-causal controller (oNCC) knows $S_t$, $Z_t^e$ as well as $Z^c_t$ before carrying out the full computation. The optimal policy is given by Lemma~\ref{lem:opt}.

\begin{lemma}\label{lem:opt}
    For each $s=(b,h)\in \mathcal{S}, b\geq u(c)$, there exists a threshold $\gamma_s$ on the confidence gain $j\triangleq z^c-z^e$ such that the policy $\pi^*(s)=g^*_s$ defined as 
    \begin{equation}\label{eq:opt_policy}
        g^*_s(z^e,z^c) = \begin{cases}
            e, & z^c-z^e\leq \gamma_s\\
            c, & z^c-z^e>\gamma_s 
        \end{cases},
    \end{equation}
    dominates any other policy $\pi\in\Pi$, that is $\rho_{\pi^*}\geq\rho_\pi$, $\forall \pi\in\Pi$.
    \begin{proof}\let\qed\relax
    See Appendix.
    \end{proof}
\end{lemma}

The optimal strategy has a threshold structure, and the computation at time $t$ is carried out across the complete DNN model only if $J_t=Z_t^c-Z_t^e > \gamma_{S_t}$. 
\begin{lemma}
    For a threshold policy $\pi$, the set of feasible Borel-measurable functions $\mathcal{G}_s$ is compact $\forall s\in\mathcal{S}$.
    \begin{proof}
        Every threshold policy $\pi=(g_{s_1},\dots,g_{s_{M}}), M=|\mathcal{S}|$ can be uniquely represented as $\gamma=(\gamma_{s_1},\dots,\gamma_{s_{M}})$, since every $g_s$ can be parameterized by $\gamma_s\in\Gamma$. It is possible to prove that there is a homeomorphism between $g_s$ and $\gamma_s$. Since compactness is a topological property, it is sufficient to show that $\Gamma$ is compact. Considering that $\mathcal{Z}$ is bounded, then $\Gamma=\text{closure}(\mathcal{Z})$ is compact.
    \end{proof}
\end{lemma}

To find the optimal values for $\gamma_s$ in each state, we need to express the expected immediate reward $\hat{r}$ and the transition probabilities $P$ of the MDP $\mathcal{M}$ as a function of $\gamma_s$. The probability of $e$ in state $s$ is
\begin{align}\label{eq:prob_exit}
    \begin{split}
        \eta(s) &= \mu_\mathbf{Z}(g^{-1}_s(e)) = \text{Pr}[Z^c-Z^e \leq \gamma_s]= F_{J}(\gamma_s),
    \end{split}
\end{align}
where $F_{J}(\cdot)$ is the cumulative distribution function of $J$. 
Hence, 
\begin{equation}\label{eq:transition_gamma}
    P(s'|s,\gamma_s) =  \text{Pr}(s'|s,e)F_J(\gamma_s) + \text{Pr}(s'|s,c)\big(1-F_J(\gamma_s)\big).
\end{equation}
For brevity, we define the decision region for action $e$ and $c$ as $\Phi^e_{\gamma_s}=\{ (z^e,z^c)\in\mathcal{Z}: z^e+\gamma_s \leq z^c \}$ , and $\Phi^c_{\gamma_s}=\mathcal{Z}\setminus \Phi^e_{\gamma_s}$, respectively. The reward $\hat{r}$ can then be expressed as 
\begin{equation}\label{eq:reward_gamma}
    \hat{r}(s,\gamma_s) = \mathbb{E}_{\mathbf{Z}}\bigg[  Z^e\mathds{1}_{\{\Phi^e_{\gamma_s}\}}(\mathbf{Z}) + Z^c\mathds{1}_{\{\Phi^c_{\gamma_s}\}}(\mathbf{Z})\bigg].
\end{equation}
Denoting $P_\gamma^\infty$ as the limiting distribution induced by $\gamma$, the optimization problem is formulated as
\begin{equation}\label{eq:gamma_opt}
    \gamma^* = \argmax_{\gamma\in\Gamma^M} \sum_{s\in\mathcal{S}}P_\gamma^\infty(s) \hat{r}(s,\gamma_s),
\end{equation}
and solved using policy iteration (PI) for average reward problems \cite{Puterman}. 

As previously mentioned, oNCC relies on the future confidence at the final exit, making it impractical for deployment. The next section addresses this issue and presents a causal, thus feasible, controller.

\subsection{Feasible Causal Controller}\label{sec:CC}
We design a causal controller (CC) to imitate the optimal behavior of the oNCC without expensive energy overhead. We equip the controller with a predictor module to estimate the conditional probability of exiting 
\begin{equation}\label{eq:cond_pexit}
    \nu(s,z) = p(z\in\Phi^e_{\gamma_s})=\text{Pr}[A_t=e|Z_t^e=z, S_t=s],
\end{equation}
$\forall z\in\mathcal{Z}^e, s=(b,h)\in\mathcal{S}, b>u(c)$. Accordingly, a stochastic policy $\hat{\pi}:\mathcal{S}\times\mathcal{Z}^e\to\mathscr{P}(\mathcal{A})$ is derived, which specifies in each state the probability distribution $q_{\hat{\pi}(s,z)}(\cdot)\in\mathscr{P}(\mathcal{A})$ over the actions.
We adopt a \emph{Gaussian naive Bayes} classifier and we address the estimation of \eqref{eq:cond_pexit} as a binary classification problem.
The set of input examples $\{z_n\}_{n=1}^N, z_n\in\mathcal{Z}^e$ are associated with the corresponding targets $\{t_n\}_{n=1}^N$, $t_n=\mathds{1}_{\{\Phi^e_{\gamma_s}\}}(z_n), \forall n$. We use the posterior class probability $p(t|z)$ learned by the model to estimate \eqref{eq:cond_pexit}. Hence, for every input $z\in\mathcal{Z}^e$, the estimation of the probability of exit is
\begin{align}
    \hat{\nu}(s,z) &= p(z\in\mathcal{Z}^e_{\gamma_s}) = p(t=1|s,z),
\end{align}
where the class conditional densites $p(z|t=i,s)$ is $\mathcal{N}(\mu_i, \sigma_i^2)$, $i=0,1$, and $\mu_i,\sigma^2_i$ are estimated with maximum likelihood. Finally, the implemented policy is defined as
\begin{equation}
    q_{\hat{\pi}(s,z)}(e) = \hat{\nu}(s,z), \quad  q_{\hat{\pi}(s,z)}(c) = 1-\hat{\nu}(s,z).
\end{equation}

\subsection{Energy-Agnostic Oracle}\label{sec:EAO}
To benchmark the long-term accuracy performance of both oNCC and CC, we design an energy-agnostic oracle-aided (EAO) controller.
At time $t$, $\forall s=(b,h)\in\mathcal{S}$, the controller chooses action: (i) $e$ if $b\geq u(e)$ and $\hat{Y}^e_t$ is correct, (ii) $c$ if $b\geq u(c)$ and $\hat{Y}^c_t$ is correct while the prediction at the early exit $ \hat{Y}^e_t$ is incorrect, otherwise (iii) an energy-free action $\varsigma$, to randomly guess the correct label. Similarly to previous controllers, when $b < u(e)$, the only feasible action is $d$.

\section{Numerical Results}\label{sec:res}
In this section, we support our theoretical analysis with numerical results for a ResNet-based~\cite{resnet} DNN architecture across CIFAR-10 \cite{cifar10} image classification dataset.\\
\indent \textbf{Overview}. First, we compute the optimal policy for oNCC using PI, and implement the feasible CC as described in Section~\ref{sec:CC}. The expressions in \eqref{eq:transition_gamma} and \eqref{eq:reward_gamma} cannot be fully computed analytically, and an empirical estimation is required. Second, to assure that the confidence of the
model is a representative estimate of the true likelihood $\text{Pr}[\hat{Y}_t^a=Y_t|A_t=a, S_t]$, a confidence calibration step is performed using temperature scaling \cite{NNcalibration}. Moreover, we compare the proposed methods, i.e., oNCC and CC, against the following benchmark controllers: (i) \emph{always continue} (A$_\ell$C), (ii) \emph{always exit} (A$_\ell$E), and (iii) the EAO controller discussed in Section.~\ref{sec:EAO}  . 
The simulation is averaged across $5$ episodes of time-horizon $T=10000$, and the following performance metrics are considered:
\begin{itemize}
    \item \emph{service rate} $\tau$: reflects the rate at which the controller provides predictions on the inputs, i.e., when $A_t\neq d$
    $$\tau=\frac{1}{T}\sum\limits_{t=0}^{T-1}\mathds{1}_{\{A_t\neq d\}}(S_t, A_t),$$

    \item \emph{accuracy} $\varrho$: captures the rate at which the controller makes correct predictions when $A_t\neq d$
    $$\varrho =\frac{1}{T\cdot\tau}\sum\limits_{t=1}^{T-1}\mathds{1}_{\{\hat{Y}_t^{A_t}=Y_t \}}(S_t, A_t),$$

    \item \emph{effective accuracy} $\alpha= \varrho\tau$: represents the overall accuracy attained in the constrained setting and measures the rate of correct predictions under energy-depletion constraints,
\end{itemize} 
where $\tau, \varrho, \alpha \in[0,1]$, $\alpha \leq \tau$ and $\alpha\leq \varrho$.\\

\begin{figure}[t]
    \centering
    \input{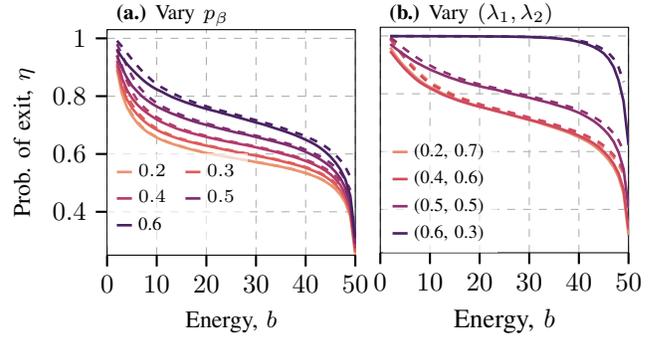}
    \caption{Learned policies over the $\mathcal{D_{\text{est}}}$, when $p_G=0.9$, and (\textbf{a}.) $\lambda_1=0.2$, $\lambda_2=0.7$, and (\textbf{b}.) $p_\beta=0.6$. Solid and dashed lines represent the policy for states $\{ G,\beta \}$, respectively.}
    \label{fig:policies}
\end{figure}
\subsection{Dataset}
The analysis consists of the following processing steps: DNN training and calibration, empirical estimation of $\hat{r}$ and $F_J$ to learn the optimal policy through PI, and the training of the naive Bayes classifier for the CC. Thus, we split the standard training partition of CIFAR-10 into four folds, $\mathcal{D}_{\text{train}}, \mathcal{D}_{\text{cali}}, \mathcal{D}_{\text{est}}, \mathcal{D}_{\text{nb}}$, with the following splitting percentage: $40\%$, $10\%$, $25\%$, $25\%$, respectively. The testing partition $\mathcal{D}_{\text{test}}$ is the standard CIFAR-10 test partition.\\
\subsection{DNN architecture}
We implement a DNN with $L=5$ residual layers \cite{resnet} $\{l_0,l_1,\dots,l_4\}$ as shown in Fig. \ref{fig:EHdevice_schema}. They consist of $\{1,6,8,12,6\}$ 3$\times$3 convolutional layers with $\{64,64,128,256,512\}$ filters, respectively. The two exit classifiers $v_0$ (early), and $v_1$ (final) are identical and consist of an average pooling, a 10-way fully-connected layer, and a softmax. They are attached to the DNN model after $l_2$ and $l_4$, respectively. The test set accuracy obtained for the early exit is $0.76$ and $0.93$ for the final exit. The number of floating point operations (FLOPs) for running the model up to $v_0$ is approximately half the FLOPs for the full DNN model. Besides, the energy cost and number of FLOPs are correlated for DNN architectures that rely on the same structure type \cite{Henderson_flops_energy}. Thus, to preserve this ratio, we assign the action costs as follows; $u(e)=1$, and $u(c)=2$.\\

\subsection{oNCC policies}%
The empirical evaluations of $\hat{r}$ and $F_j$ are computed over $\mathcal{D_{\text{est}}}$: the former through a Monte-Carlo estimation, while the latter using the empirical cumulative distribution.
The optimal policy $\gamma^*$ for oNCC is obtained using PI. Given the one-to-one mapping between $\gamma$ and the probability of exit $\eta$ in \eqref{eq:prob_exit}, we discuss the obtained policy for oNCC in terms of $\eta$. 
In Fig.~\ref{fig:policies}, we show the effect of the current energy $b\in\mathcal{B}$, the EH state $h\in\{G,\beta\}$, and the EH rates on the behaviour of the oNCC policy. In Fig.~\ref{fig:policies}.a, we vary $p_B$ while in Fig.~\ref{fig:policies}.b, we vary $\lambda_i$ to simulate various EH rate combinations. Overall, the probability of exit $\eta$ increases as the battery depletes, $B_t\to0$, resulting in more energy-conservative policies. In fact, the controller is forced to \emph{discard} the sample when the battery is depleted, thus obtaining zero reward. On the other hand, when $B_t\to b_{max}$, the controller prevents energy buffer overflows by decreasing the probability of exit $\eta$, resulting in a more permissive strategy.
Moreover, the policy for \emph{bad} states (dashed) is marginally more conservative than the one for \emph{good} states (solid), especially when $p_\beta$ is high. Besides, in Fig.~\ref{fig:policies}.b, the two configurations $(\lambda_1,\lambda_2)=(0.2,0.7)$, and $(\lambda_1,\lambda_2)=(0.4,0.6)$ have the same average EH rate $\varepsilon = 1.28$ and induce the same probability of exit. However, in general, the EH rate is not enough to determine the optimal exit policy, as $p_G$ and $p_\beta$ also play a crucial role in shaping the policy. Interestingly, as shown in Fig.~\ref{fig:policies}.b with configuration $(\lambda_1,\lambda_2)=(0.6,0.3)$, even when the EH rate is too low, i.e., $\varepsilon < u(e)$, the learned policy tends to decrease the exiting probability as $B_t \to b_{max}$.

\subsection{Controllers comparison}
\begin{figure}[t]
    \centering
 
 
 
\begin{tikzpicture}

\definecolor{indigo}{RGB}{74,34,98}
\definecolor{salmon}{RGB}{231,125,99}
\definecolor{steelblue}{RGB}{31,119,180}
\definecolor{indianred}{RGB}{208,73,97}
\definecolor{darkgray}{RGB}{176,176,176}

\begin{axis} [
    ybar = .06cm,
    name=perf,
    bar width = 6pt,
    width=7cm, height=3.5cm,
    cycle list ={indianred, salmon, indigo},
    ymin = 0.55, 
    ymax = 1., 
    symbolic x coords={A$_\ell$C, A$_\ell$E, EAO, oNCC, CC},
    legend style = {
        at={(0.5, 0.)}, 
        anchor=south,
        legend columns=-1,
        fill opacity=0.8,
        text opacity =1,
    },
    enlarge y limits = {abs = .07, upper},
    y grid style={darkgray},
    ymajorgrids=true,
    xtick align=outside,
    ytick align=inside,
    xtick pos=left,
    ytick pos=left,
    xtick={A$_\ell$C, A$_\ell$E, EAO, oNCC, CC},
    ylabel={$\alpha$, $\varrho$, $\tau$},
    nodes near coords,
    nodes near coords style={
            font=\scriptsize,
            rotate=90,
            anchor=west,
        },
    every axis plot/.append style={
        fill,
        fill opacity=.8,
        draw opacity=1,
        text opacity=1,
    }
]
 
\addplot table[x=x, y=mean, col sep=comma] {data/alpha_mean_performance_noRND.csv};

\addplot table[x=x, y=mean, col sep=comma] {data/rho_mean_performance_noRND.csv};

\addplot table[x=x, y=mean, col sep=comma] {data/tau_mean_performance_noRND.csv};

\legend {$\alpha$, $\varrho$, $\tau$ };
 
\end{axis}
 
\end{tikzpicture}
 
    \caption{ Effective accuracy $\alpha$, accuracy $\varrho$, and service rate $\tau$ computed over the test set $\mathcal{D}_{\text{test}}$, when $\lambda_1=0.2$, $\lambda_2=0.7$, $p_G= 0.9$, $p_\beta=0.6$, $\varepsilon=1.28$, $T=10^4$, $b_{max}=50$.}
    \label{fig:barplot}
\end{figure}
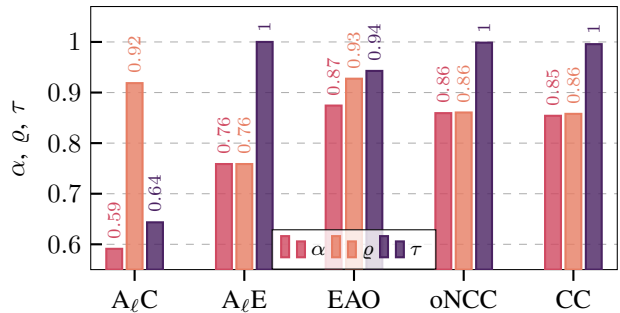

\begin{figure*}[t]
    \centering
    \input{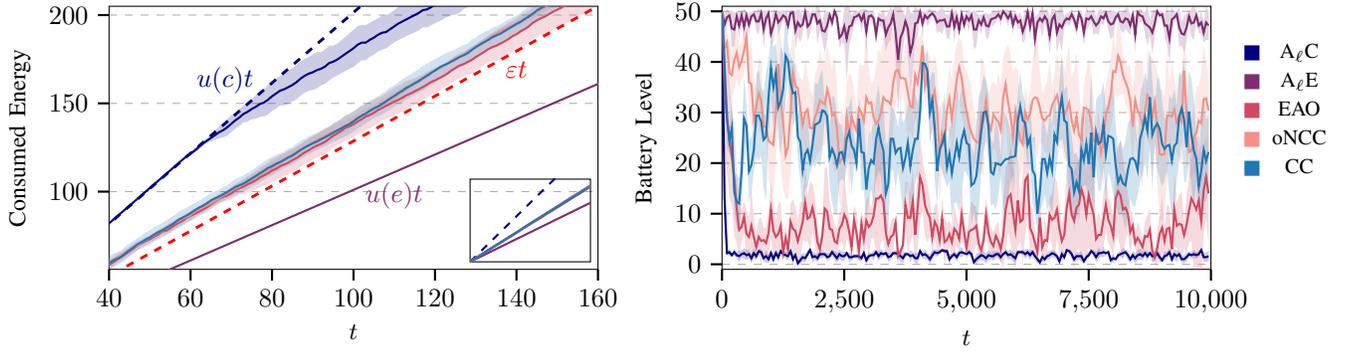}
    \caption{ Mean and confidence intervals of cumulative consumed energy (left) and battery level (right), with the configuration setting of Fig.~\ref{fig:barplot}. The inset graph (left) shows the behavior of the consumed energy in long-term horizon.}
  \label{fig:energy_plot}
\end{figure*}

We benchmark the proposed methods over $\mathcal{D}_{\text{test}}$ in terms of service rate $\tau$, accuracy $\varrho$, and effective accuracy $\alpha$, to analyze the performance of the controllers from different perspectives. 
We discuss the controllers' behavior with respect to the numerical results reported in Fig.~\ref{fig:barplot}.
As expected, the A$_\ell$C controller attains the full model's accuracy \mbox{$\varrho\approx 0.92$}. However, since the energy cost 
of \emph{continue}, $u(c)=2$, is higher than the energy rate $\varepsilon=1.28$, a quick depletion of the battery is observed in Fig.~\ref{fig:energy_plot}. The service rate $\tau$ and effective accuracy $\alpha$  are impacted considerably and  drop to approximately $0.6$. Consequently, deploying a DNN model in an energy-constrained scenario without any energy-adaptive measures impacts the effective accuracy $\alpha$ by $\approx 40\% $ w.r.t. to $\varrho$.
The implemented early exit allows to reduce the energy cost of a prediction. Since in this setting the rate of consumed energy at the early exit, $u(e)=1$ is lower than $\varepsilon=1.28$, the A$_\ell$E controller is very unlikely to deplete the battery and achieves a service rate very close to $\tau=1$. As a consequence, $\varrho=\alpha\approx0.76$, which is the accuracy obtained at the early exit. Hence, the performance of the system is limited due to the non-efficient exploitation of the incoming energy, leading to battery overcharge.\\
The EAO controller represents a non-feasible sample-sensitive combination of A$_\ell$E and A$_\ell$C. It behaves greedily towards accuracy and attains the maximum achievable accuracy $\varrho\approx 0.93$, which is marginally higher than A$_\ell$C due to DNN overthinking. However, due to its energy-agnostic nature, the service rate $\tau$ and effective accuracy $\alpha$ decrease by $6\%$ in our simulated scenario. Under severe energy-constrained scenarios, or a larger accuracy gap between A$_\ell$E and A$_\ell$C, the performance reduction can become significant.\\
\indent The proposed policy implements an energy-aware strategy to intelligently control the computation pipeline of the DNN model. Even with partial knowledge of the side-information vector, the proposed naive Bayes predictor enables the feasible CC to accurately match the performance of oNCC with respect to all three metrics. Fig.~\ref{fig:barplot} shows that the developed energy-aware strategies achieve continuous availability, $\tau\approx 1$. Conversely to A$_\ell$E, the energy consumption rate of the proposed controllers matches the incoming energy rate as illustrated in Fig.~\ref{fig:energy_plot}. Hence, the system pushes towards an efficient energy exploitation. As a consequence, the proposed method improves the accuracy by $12\%$ w.r.t. A$_\ell$E, indicating that the CC adapts to its environment and minimizes the impact of the limited energy. In fact, the effective accuracy of CC reaches the performance of both the EAO controller and oNCC.

\section{Conclusion}\label{sec:conc}
In this work, we addressed the challenges of limited and stochastic energy availability in IoT edge devices, to enable a sustainable processing paradigm for  edge intelligence.
We considered a time-correlated energy source and a finite battery for storage. We adopted EE as a method to dynamically adapt the energy consumption of a DNN model. Formulating the problem as MDP-SI, we proved that, for a given energy setting, the optimal EE policy has a threshold structure, which depends on the confidence gain between the early and final exits. Using a naive Bayes predictor, we developed a causal controller that mimics the behaviour of the optimal, yet non-feasible, non-causal controller. To complement our theoretical analysis, we provided numerical results on image classification with different environment parameters. We showed that the implementation of EE coupled with an intelligent exit policy is essential for reliable edge intelligence under limited and/or stochastic energy availability. Future work will extend this study to encompass multiple early exits.

\appendix[Proof of Lemma \ref{lem:opt}]
\begin{proof}
    In each state $s$, a stationary policy $\pi(s)=g_s$ induces a probability over actions, $\mathscr{P}_\pi(\mathcal{A})$,  that is $p_{\pi(s)}(a)=\mu_{\mathbf{Z}}(g_s^{-1}(a)), a\in\mathcal{A}$. We define, for each $\pi\in\Pi$, the class of policies that induce the same probability $\mathscr{P}_\pi(\mathcal{A})$ as $\pi$, i.e., $\Lambda_\pi=\{\pi'\in\Pi: \mathscr{P}_{\pi'}(\mathcal{A})=\mathscr{P}_{\pi}(\mathcal{A})\}$. Since the MDP has a finite state space and the reward is bounded, then the limiting distribution induced by a stationary policy $\pi$, $P_\pi^\infty$, exists and is stochastic \cite{Puterman}. Therefore, 
    \begin{equation}
        \rho_\pi(s) = \sum_{s'\in\mathcal{S}}P_\pi^\infty(s'|s)\,\hat{r}(s',\pi(s')),\quad \forall s \in \mathcal{S}.
    \end{equation}
    The best policy $\pi^*$ in $\Lambda_\pi$ satisfies $\rho_{\pi^*}(s)\geq\rho_{\pi'}(s), \forall s\in\mathcal{S}, \pi'\in\Lambda_\pi$, hence
    \begin{equation}
        \rho_{\pi^*}(s) = \max_{\pi'\in\Lambda_{\pi}} \sum_{s'\in\mathcal{S}}P_{\pi'}^\infty(s'|s) \hat{r}(s',\pi'(s'))
    \end{equation}
    Since policies in $\Lambda_\pi$ have the same transition dynamics, thus the same limiting distribution $P_\pi^\infty$, the optimal policy $\pi^*$ is the one that maximizes the expected immediate reward subject to $\pi^*\in\Lambda_\pi$, i.e.,
    \begin{argmaxi!}|s|
        {\pi'\in\Pi}{ \hat{r}(s, \pi'(s))}
        {}{\pi^*(s) =}
        \addConstraint{\mu_{\mathbf{Z}}\big({\pi'(s)}^{-1}(e)\big)=p_e},
    \end{argmaxi!}
    where $p_e\triangleq p_{\pi(s)}(e)$.
    We form the Lagrangian function 
    \begin{align*}
        \begin{split}
        L(\pi'_s, \gamma_s) &= \mathbb{E}_{\mathbf{Z}}\bigg[  Z^e\mathds{1}_{\{{\pi'_s}^{-1}(e)\}}(\mathbf{Z}) + Z^c\mathds{1}_{\{{\pi'_s}^{-1}(c)\}}(\mathbf{Z})\bigg]\\
            &+\gamma_s\,\big(\mathbb{E}_{\mathbf{Z}}[\mathds{1}_{\{{\pi'_s}^{-1}(e)\}}(\mathbf{Z})] - p_e\big),
        \end{split}
    \end{align*}
    where we abbreviate $\pi(s)$ as $\pi_s$, and solve the Lagrangian unconstrained problem, obtaining
     \begin{equation}\label{eq:opt_lagrangian_sol}
            \mathds{1}_{\{\pi_s^{*^{-1}}(e)\}}(z^e,z^c)=\begin{cases}
                1 & z^e+\gamma_s \geq z^c\\
                0 & z^e+\gamma_s < z^c.
            \end{cases}
        \end{equation}
        Due to the continuity of $\mu_{\mathbf{Z}}$ it is possible to prove that, given a policy $\pi$ and its transition dynamics equivalence class $\Lambda_\pi$, $\Lambda_\pi$ always contains a policy $\pi^*$ with the structure described in \eqref{eq:opt_lagrangian_sol}. In other words, the policy structure in \eqref{eq:opt_lagrangian_sol} is feasible for the problem, thus for the Lagrangian sufficiency theorem is optimal.
\end{proof}


\bibliographystyle{IEEEtran}
\bibliography{IEEEabrv,sust_edge_intelligence_through_EAEE.bib}

\end{document}